\documentclass[12pt,twoside]{article}
\usepackage{fleqn,espcrc1}

\usepackage{graphicx}
\usepackage[figuresright]{rotating}

\def\bra{\,<\!} \def\ket{\!>\,} \def\ack{\,|\,}

\title{Nuclear magnetic dipole properties and the triaxial deformation}

\author{
Yang Sun, \address{Department of Physics and Astronomy, University of
Tennessee, Knoxville, Tennessee 37996, USA}
\address{Department of Physics, Xuzhou Normal University,
Xuzhou, Jiangsu 221009, P.R. China}
Javid A. Sheikh, \address{Physik-Department, Technische
Universit\"at M\"unchen, D-85747 Garching, Germany}
Gui-Lu Long \address{Department of Physics, Tsinghua University, Beijing
100084, P.R. China}
}

\begin{document}
\maketitle

\bigskip
\begin{abstract}

ABSTRACT: Nuclear magnetic dipole properties of ground bands and
$\gamma$-vibrational bands are studied for the first time using
the triaxial projected shell model approach. The study is carried out
for the Dy and Er isotopic chains, ranging from transitional to
well-deformed region. It is found that the g-factor ratio of the $2^+$ state in
ground bands to that of $\gamma$-bands, $r=g(2^+, 
\gamma$-vib)/$g(2^+,$ ground), varies along an isotopic chain.  
With the $\gamma$-deformations, which best reproduce
the energy levels for both bands, we obtain a  
qualitative agreement with the experimental data. This result thus suggests 
that study of the ratio may provide an important information on the
triaxial deformation of a nuclear system.
The angular-momentum dependence of the ground band g-factor on the triaxial deformation
is also investigated.
\end{abstract}

\bigskip
\bigskip

The importance of triaxial deformation has been a long-standing problem 
in the studies
of atomic nuclei. Although it is known from several model studies that nuclei in
the transitional regions acquire a stable triaxial mean-field deformation, 
consequences of it on the measurable electromagnetic transition probabilities 
have remained less clear. 
Nuclear magnetic dipole properties are very sensitive to the
single-particle aspects of the nuclear wave-function and hence can
provide valuable information on the microscopic structure of a
nuclear system. As the single-particle orbits can be strongly modified by
triaxiality, the interesting question is whether the triaxial deformation
will have observable effects in the magnetic dipole properties.

The study of nuclear magnetic properties
has been an active field of research in nuclear structure physics. 
On the experimental side, 
considerable progress in the g-factor experiment 
has recently been made by taking advantages of modern
experimental techniques and sensitive detectors, for a recent
example, see Ref. \cite{Ke01}. 
There have been new measurements 
on the ground band and the $\gamma$-band g-factors for
some rare earth nuclei \cite{g-Dy,Bonn96}. 
On the theoretical side, 
several early calculations
were devoted to the study of g-factors. They were based either on the
cranking models \cite{BA86,TT88,Ces91,SS93} or on the angular
momentum projection method \cite{An90,Sun94,Vel99}. There have been
also calculations using the interacting boson model \cite{KS95}.
However, a microscopic and systematic investigation that treats ground bands
and $\gamma$-bands simultaneously, and thus can study their
correlations along an isotopic chain, is still missing.

Recently, triaxial projected shell model (TPSM) calculation
has become available \cite{SH99}. The TPSM extends the original
projected shell model \cite{review} by removing the restriction
of an axially deformed basis assumed in the earlier calculations. In the
TPSM approach, one introduces triaxiality in the deformed basis
and performs exactly three-dimensional angular momentum
projection. In this way, the deformed vacuum state is much
enriched by allowing all possible $K$-components. Diagonalization
mixes these $K$-states, and various excited levels, specified by
$K$, emerge \cite{Sun00}. The first excited TPSM band describes
the observed $\gamma$-vibrational band, and the second excited
TPSM band reproduces the experimental $\gamma\gamma$-band. This
picture is similar to the one shown by Davydov and Filippov
\cite{DF58}, but now obtained in terms of a fully microscopic
theory. Transition quadrupole moments in $\gamma$-soft nuclei
have also been well described by the TPSM \cite{SSP01}.
It was shown \cite{SSP01} that the increasing trend in 
transition quadrupole moment of ground 
bands has a close relation to the triaxiality of a nucleus. 

In the present work, the ground band and the $\gamma$-band g-factors
are studied for the first time by using the TPSM. 
As in the early projected shell model studies \cite{review},
to treat these deformed rare earth nuclei, it is appropriate to
include three major shells for each type of
nucleons ($N=4,5,6$ for neutrons and $N=3,4,5$ for protons). 
The TPSM wave-function can be written as
\begin{equation}
\Psi^{\sigma I}_{M} = \sum _{K} f^{\sigma I}_{K}\,\hat
P^I_{MK}\ack\Phi\ket .
\label{wf}
\end{equation}
In Eq. (\ref{wf}), $\sigma$ specifies the states with the same angular momentum
$I$, $\hat P^I_{MK}$ is the three-dimensional projection operator
\begin{equation}
\hat P^I_{MK} =
{2I+1 \over 8\pi^2} \int d\Omega\, D^{I}_{MK}(\Omega)\, \hat R(\Omega),
\end{equation}
and $\ack\Phi\ket$ represents the triaxial qp vacuum state 
\begin{equation}
\left\{\hat P^I_{MK}\ack\Phi\ket,~0 \le K \le I \right\}.
\label{basis}
\end{equation}
For studying the low-spin states with $0\le I\le 10$ before
any quasi-particle (qp) alignment may occur, this is
the simplest possible configuration space for an even-even nucleus 
with the $\gamma$-degree of freedom included.

As in the earlier calculations, 
we use the pairing plus
quadrupole-quadrupole Hamiltonian 
with inclusion of the quadrupole-pairing term 
\begin{equation}
\hat H = \hat H_0 - {1 \over 2} \chi \sum_\mu \hat Q^\dagger_\mu
\hat Q^{}_\mu - G_M \hat P^\dagger \hat P - G_Q \sum_\mu \hat
P^\dagger_\mu\hat P^{}_\mu .
\label{hamham}
\end{equation}
The corresponding triaxial Nilsson Hamiltonian is given by
\begin{equation}
\hat H_N = \hat H_0 - {2 \over 3}\hbar\omega\left\{\epsilon\hat Q_0
+\epsilon'{{\hat Q_{+2}+\hat Q_{-2}}\over\sqrt{2}}\right\}.
\label{nilsson}
\end{equation}
In Eq. (\ref{hamham}), $\hat H_0$ is the spherical
single-particle Hamiltonian, which contains a proper spin--orbit
force \cite{NKM}. The interaction strengths are taken as follows:
The $QQ$-force strength $\chi$ is adjusted such that it has a
self-consistent relation with the quadrupole deformation
$\epsilon$. 
The monopole pairing strength $G_M$ is
of the standard form $G_M = \left[21.24
\mp13.86(N-Z)/A\right]/A$, with ``$-$" for neutrons and ``$+$"
for protons, which approximately reproduces the observed
odd--even mass differences in this mass region. 
The quadrupole pairing strength $G_Q$ is assumed to be
proportional to $G_M$, the proportionality constant being fixed
as usual to be 0.16. These interaction
strengths are consistent with those used previously for the same
mass region \cite{SH99,review,Sun00,SSP01}.
$\epsilon'$ in Eq. (\ref{nilsson}) can be adjusted, reflecting how much the
system is $\gamma$-deformed.
The philosophy
of the present approach is that based on the Nilsson potential,  
one performs explicit angular-momentum projection 
with a two-body interaction which conforms (through self-consistent
conditions) with the mean-field Nilsson potential.
The Hamiltonian with separable forces 
serves only as an effective interaction,
the parameters of which have been fitted to the experimental data.

The Hamiltonian in Eq. (\ref{hamham}) is diagonalized using the projected basis
of Eq. (\ref{basis}). 
We emphasize that, although only the qp-vacuum state is included in the basis, 
its triaxial nature generates the
$K$-mixing when the diagonalization is carried out.
The expansion coefficients $f$, obtained through the
diagonalization of the shell-model Hamiltonian,
describe the amount of $K$-mixing
and specify various physical states 
(e.g. g-, $\gamma$-, $\gamma\gamma$-bands, $\cdots$)
\cite{Sun00}.

\begin{figure}[htb]
\begin{minipage}[t]{150mm}
\includegraphics[scale=0.50]{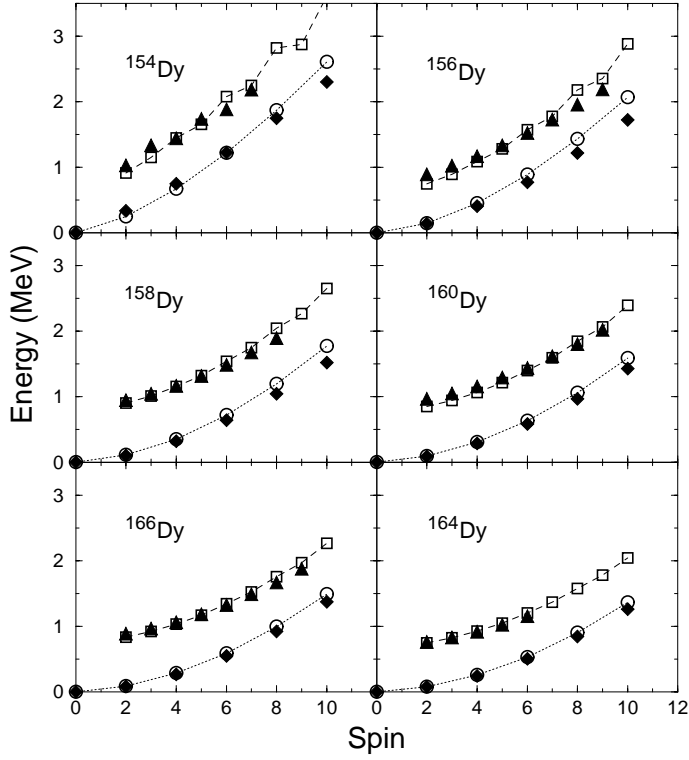}
\caption{Comparison of calculated energy levels with 
experimental data for the ground bands
and the $\gamma$-band in $^{154-164}$Dy.
Filled symbols represent data and open symbols the calculation. 
}
\label{fig1}
\end{minipage}
\end{figure}

\begin{table}
\caption{Deformation parameters $\epsilon$ and $\epsilon'$ used in the 
TPSM calculations. $\gamma$ are obtained from the relation 
$\tan \gamma = {\epsilon' \over \epsilon}$.} 
\begin{tabular}{|cccc|}
\hline
Nucleus & $\epsilon$ & $\epsilon'$ & $\gamma$ (degree)\\
\hline
$^{154}$Dy & 0.21 & 0.120 & 30\\
$^{156}$Dy & 0.24 & 0.130 & 28\\
$^{158}$Dy & 0.26 & 0.120 & 25\\
$^{160}$Dy & 0.27 & 0.125 & 25\\
$^{162}$Dy & 0.28 & 0.130 & 25\\
$^{164}$Dy & 0.29 & 0.140 & 26\\
\hline
\end{tabular}
\label{table1}
\end{table}

In Fig. 1, we present the calculated energy levels 
together with experimental data for the ground and
the $\gamma$-band in $^{154-164}$Dy. 
In Table 1, $\epsilon$ and $\epsilon'$ used in the 
TPSM calculations for the Dy isotopes are listed. 
It can be seen that the TPSM describe the energy levels very well.
In particular, the steepness in the increasing curve for the lighter isotopes
and the position of the $\gamma$-band head 
are correctly described 
in the calculation.  
In the later discussion for the g-factor calculations, the same deformations
$\epsilon$ and $\epsilon'$ in Table 1 
will be used. We note that changing $\epsilon'$ can give different
g-factor results, but this will destroy the agreement for the
spectrum calculation obtained in Fig. 1.

The wave-functions obtained from the diagonalization are 
then used to evaluate the electromagnetic
transition probabilities.
The g-factor $ g(\sigma, I)$ is 
generally defined as
\begin{equation}
g(\sigma, I) = \frac {\mu(\sigma, I)}{\mu_N I} = g_\pi (\sigma,I) + g_\nu
(\sigma,I) ,
\end{equation}
with $\mu(\sigma, I)$ being the magnetic moment of a state $(\sigma, I)$.
$g_\tau (\sigma, I), \tau = \pi$ or $\nu$, is given by
\begin{eqnarray*}
g_\tau (\sigma, I) &=& {1\over{\mu_N I}} \bra \Psi^{\sigma}_{II} |
\hat \mu^\tau_z | \Psi^{\sigma}_{II} \ket \nonumber \\
&=& {1\over{\mu_N \sqrt{I(I+1)}}} \bra \Psi^{\sigma}_{I} ||
\hat \mu^\tau || \Psi^{\sigma}_{I} \ket \\
&=& \frac{1}{\mu_N \sqrt{I(I+1)}} \left(
     g^{\tau}_l \bra \Psi^\sigma_I||\hat j^\tau ||\Psi^\sigma_I \ket
     + (g^{\tau}_s - g^{\tau}_l) 
     \bra\Psi^\sigma_I||\hat s^\tau||\Psi^\sigma_I\ket \right) .
\end{eqnarray*}
In our calculations, the following standard values for $g_l$ and
$g_s$ \cite{BM75} have been taken: $ g_l^\pi = 
1, g_l^\nu = 0, g_s^\pi = 5.586 \times 0.75 $, and $ g_s^\nu =
-3.826 \times 0.75 $.
In the angular-momentum projection theory, 
the reduced matrix element for $\hat m$ (with $\hat m$ being either 
$\hat j$ or $\hat s$) can be explicitly expressed as
\begin{eqnarray*}
&\bra& \Psi^{\sigma}_{I} || \hat m^\tau ||
\Psi^{\sigma}_{I} \ket
\nonumber \\
&=& \sum_{K_i,K_f} f_{I K_i}^{\sigma} f_{I K_f}^{\sigma}
\sum_{M_i , M_f , M} (-)^{I - M_f}
\left(
\begin{array}{ccc}
I & 1 & I \\-M_f & M &M_i
\end{array} \right)
\bra \Phi | {\hat{P}^{I}}_{K_f M_f} \hat m_{1M}
\hat{P}^{I}_{K_i M_i} | \Phi \ket\nonumber \\
 &=& (2I+1) \sum_{K_i,K_f} (-)^{I-K_f} f_{I K_i}^{\sigma} f_{I K_f}^{\sigma}
\nonumber \\
& & \times \sum_{M^\prime,M^{\prime\prime}}
\left(
\begin{array}{ccc}
I & 1 & I \\-K_{f} & M^\prime & M^{\prime\prime}
\end{array} \right)
\int d\Omega {\it D}_{M'' K_{i}} (\Omega)
\bra \Phi | \hat m_{1M'} \hat{R}(\Omega) | \Phi \ket .
\end{eqnarray*}
We finally have
\begin{eqnarray*}
g_\tau (\sigma, I) &=& {1\over{\mu_N (I+1)}}
\sum_{K_i,K_f} f_{I K_i}^{\sigma} f_{I K_f}^{\sigma}
\sum_{M^\prime,M^{\prime\prime}} \bra I M^{\prime\prime} 1 M^\prime
 | I K_f \ket \nonumber \\
&& \times \int d\Omega {\it D}_{M'' K_{i}} (\Omega)
\bra \Phi | \hat m_{1M'} \hat{R}(\Omega) | \Phi \ket .
\end{eqnarray*}

Since wave functions for a ground band and a $\gamma$-band contain 
different components of the $K$-states, they can have
different responses to the triaxiality. As a
consequence, one expects different behavior of the g-factors of the
states in a ground band and a $\gamma$-band as a function
of $\gamma$-deformation. An interesting relation between the ground
band and the $\gamma$-band g-factors has been suggested based on
experimental information \cite{Bonn96}, and the departure of the ratio $r=g(2^+,
\gamma$-vib)/$g(2^+,$ ground)
from 1 was interpreted as the
$F$-spin admixture within the interacting boson model \cite{GF92}.

\begin{figure}[htb]
\begin{minipage}[t]{75mm}
\includegraphics[scale=0.50]{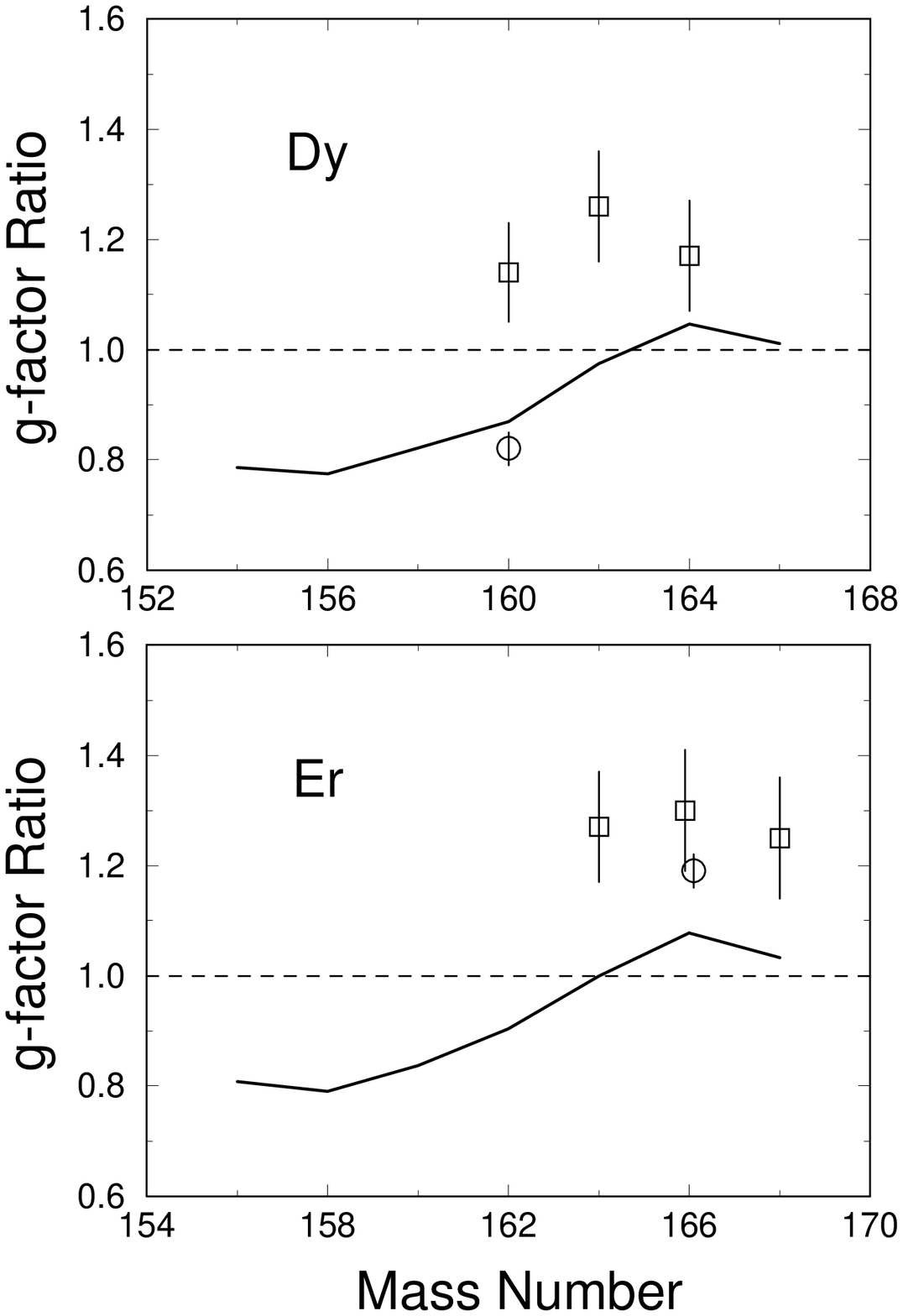}
\caption{Comparison of calculated g-factor ratio $r=g(2^+,
\gamma$-vib)/$g(2^+,$ ground) with available data for Dy (top panel)
and Er (bottom panel) isotopes. 
Data are taken from Refs. \protect\cite{g-Dy,Bonn96,Bonn95}.
}
\label{fig2}
\end{minipage}
\hspace{\fill}
\begin{minipage}[t]{75mm}
\includegraphics[scale=0.50]{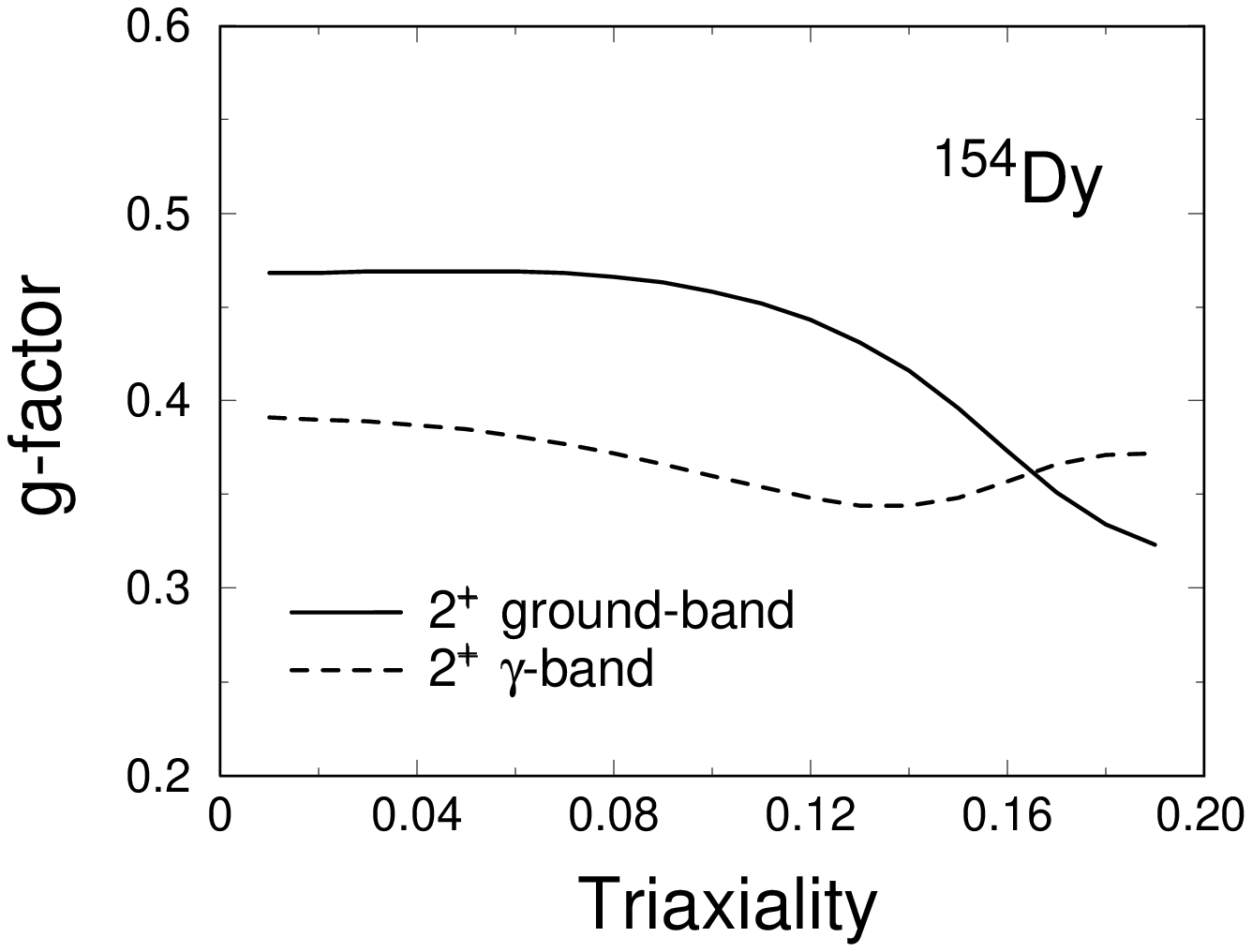}
\caption{Calculated g-factors of the $2^+$ state in the ground
band (solid line) and that in the $\gamma$-band (dashed line) for
$^{154}$Dy
as a function of triaxial
deformation $\epsilon'$.}
\label{fig3}
\end{minipage}
\end{figure}

In Fig. 2, the calculated g-factor ratios of the $2^+$
$\gamma$-state to that of the $2^+$ state of ground band 
are plotted for the Dy and Er
isotopic chains. In going from the neutron number $N=88$ to 100, 
the theoretical curves show the same trend for the two isotopic chains:
As the neutron number increases, the ratio $r$ starts from a value around 0.8,  
goes up and crosses the $r=1$ line, reaches a value close to 1.1, 
and then turns back. 
Thus, our theoretical $r$ is clearly smaller than 1 for 
transitional nuclei, and is close to or larger than 1 for 
well-deformed nuclei. 
By collecting available data from the rare earth region, 
Alfter {\it et al.} obtained a similar 
trend for the experimental $r$ \cite{Bonn96}. 
Unfortunately, there are not
sufficient data across the whole isotopic chains for a comprehensive study.  
In Fig. 2, we plot the available data of the Dy and Er isotopes. 
For nuclei with the neutron number larger than 96, we have obtained  
an $r$ value larger than 1, indicating that for these nuclei,
the g-factor of the $2^+$ $\gamma$-state is larger than that of
the $2^+$ state of ground band. This is in a qualitative 
agreement with the existing data. However,
the amplitude of the 
theoretical ratios are smaller than the experimental values. 
For the nucleus $^{160}$Dy, there have been two independent measurements
with rather controversial conclusions 
\cite{g-Dy,Bonn95}. The calculated $r$ value for $^{160}$Dy is close
to the one obtained by Alfter {\it et al.} in Ref. \cite{Bonn95}. 
The implication
of these results is discussed below.  

For a better understanding of the above results, let us study
the g-factors in terms of the triaxiality in the deformed Nilsson basis
and the $K$-mixing among the basis states. 
In Fig. 3, we plot the g-factors of the $2^+$ states in
the ground band and the $\gamma$-band for $^{154}$Dy
as a function of the triaxial
deformation parameter $\epsilon'$. 
It is seen from Fig. 3, that the
g-factor of the ground band $2^+$ state is significantly larger
than that of the $\gamma$-band if the triaxiality $\epsilon'$ is small. 
However, as $\epsilon'$ increases, the two g-factors attempt to be close to
each other. The two g-factor curves cross eventually at an higher 
triaxiality around $\epsilon' = 0.16$. 
The tendency that the two g-factors get closer for larger $\epsilon'$
may be understood as a consequence of the $K$-mixing 
among the projected $K$-states in Eq. (\ref{basis}). 
For a small $\epsilon'$, the ground band consists of a pure
$K=0$ and the $\gamma$-band a pure $K=2$ component.  
With increasing $\epsilon'$, considerable $K=2$ component is mixed
into the ground band, and {\it vice versa}, 
$K=0$ component into the $\gamma$-band 
(see also the discussion in Ref. \cite{Sun00}).
For a sufficiently large $\epsilon'$, both components can be nearly equally
present in the ground band and in the $\gamma$-band. 
This is the reason to obtain the close values for the two g-factors
with large triaxiality. 

It becomes clear from Fig. 3 that in our calculation, 
whether $r$ is larger or smaller than 1, 
as well as its amplitude, is closely 
related to the triaxiality $\epsilon'$.
Changing $\epsilon'$ can lead to a different theoretical picture.
However, the theoretical $r$ given in Fig. 2 
are obtained with the $\epsilon'$ that can best
reproduce the energy levels in Fig. 1 (for the Er isotopes, in Ref.
\cite{Sun00}). 
$\epsilon'$
is related to the conventional triaxial parameter
$\gamma$ through the relation $\tan \gamma = {\epsilon' \over
\epsilon}$. The value of $\epsilon$ is held fixed for each
nucleus in the calculations, and therefore, $\gamma$ increases
linearly with $\epsilon'$. 
We thus suggest that accurate measurement of $r$ can provide useful information
for understanding the triaxiality of a nucleus.  
Experimental study of g-factors is a very challenging task.
It is not surprising to see very
controversial results from different measurements, such as 
the example of $^{160}$Dy. 
We hope that our results will stimulate future g-factor experiments.

Now we turn our discussion to the angular-momentum dependence of
the ground band g-factors. Experimental data in Ref.
\cite{Bonn97} exhibited a significant variation in the g-factors
of the low-spin states in some Dy ground bands. The data shows a
large drop at $6\hbar$ in $^{158}$Dy and $^{160}$Dy, and a clear
jump at $8\hbar$ in $^{162}$Dy. The early projected shell model
calculations \cite{Sun94,Vel99} with an axially deformed basis
did not reproduce the variations. We note that these data are 
different from those obtained  
by Brandolini {\it et al.} \cite{g-Dy,g-Er} 
in which these variations were not seen. 
In the early projected shell model calculations, while the band-crossing
with an $i_{13/2}$ neutron pair can be excluded as a cause for
these variations \cite{Vel99}, there could exist uncertainties
due to the restriction of an axially deformed basis. 
Thus, with the new three-dimensional projection calculation 
available, the projected shell model results can now be recalculated. 

\begin{figure}[htb]
\begin{minipage}[t]{150mm}
\includegraphics[scale=0.50]{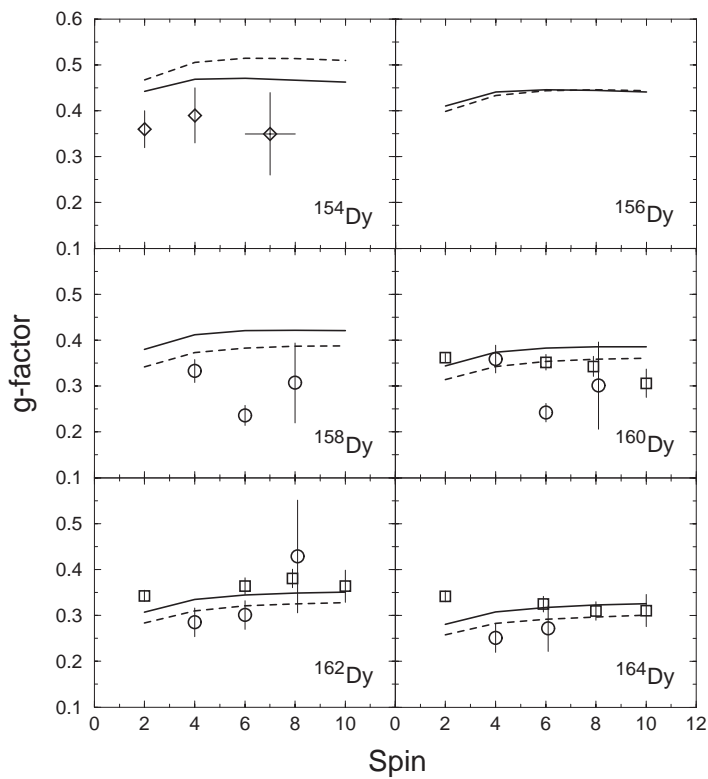}
\caption{Comparison of available data with calculated g-factors
for $^{154-164}$Dy. There are two sets of theoretical curves for
calculations with triaxially deformed basis (solid curves) and with
axially deformed basis ($\epsilon'=0$, dashed curves). The
experimental data are taken from Ref. \protect\cite{g-Dy} (open
squares), Ref. \protect\cite{Bonn97} (open circles), and Ref.
\protect\cite{g-Dy154} (open diamonds).}
\label{fig4}
\end{minipage}
\end{figure}

In Fig. 4, results of the TPSM calculations are compared with the
experimental data \cite{g-Dy,Bonn97,g-Dy154} for the isotopes
$^{154-164}$Dy. The deformations $\epsilon$ and $\epsilon'$ used
in the calculations are again those listed in Table 1. 
It can be seen from Fig. 4
that the theoretical curves appear to be smooth as a function of
angular momentum and no variation in the g-factors is obtained.
Consequently, the experimental g-factors 
by Brandolini {\it et al.} \cite{g-Dy} are nicely reproduced. 
We can thus conclude that the present
calculations, performed with inclusion of $\gamma$-deformed
basis and three-dimensional angular momentum projection,
can not produce g-factor variations in the low-spin states of
ground bands. 

In conclusion, the low-spin g-factors of both ground bands and
$\gamma$-vibrational bands in the $\gamma$-soft and
well-deformed rare earth nuclei have been studied simultaneously using the
triaxial projected shell model approach. The 
diagonalization is carried out in a shell model basis with three-dimensional
angular-momentum projection on the triaxially deformed
Nilsson-states. It has been shown that the $2^+$ state g-factors
in ground bands and in $\gamma$-bands behave differently
as a function of triaxiality. 
The g-factor ratio $r$ 
varies along an isotopic chain, having different characteristic 
values for transitional and well-deformed nuclei. 
For a given nucleus, the calculated $r$ can be changed with triaxial
deformation. 
Therefore, we suggest that the ratio of the g-factors, 
not their dependence on rotation, may be
taken as a fingerprint of the nuclear triaxiality. 
More accurate measurements of the g-factor ratio will be very helpful for a 
better comparison with the theoretical results presented in this
manuscript.

YS thanks the Department of Physics of Tsinghua University for
warm hospitality, and for support through the senior visiting
scholar program of Tsinghua University.
GLL is supported by National Natural Science
Foundation of China under Grant No. 10047001, 
and by China Major State Basic Research
Development Program under Contract
No. G200077400.

\baselineskip = 14pt
\bibliographystyle{unsrt}

\end{document}